%% file: ms.tex
\documentclass{article}
\usepackage{amsmath,graphicx,mlspconf}

\usepackage[english]{babel}    
\usepackage{amssymb}           
\usepackage[utf8]{inputenc}    
\usepackage[T1]{fontenc}       
\usepackage{eucal}             
\usepackage{bbm}               
\usepackage{longtable}         
\usepackage{exscale}           
\usepackage{mathtools}         
\usepackage[amsmath,thmmarks,hyperref]{ntheorem} 
\usepackage{graphicx}   
\usepackage[sort]{cite}        
\usepackage{mathrsfs}          
\usepackage{array}             
\usepackage{stmaryrd}          

\input{DSBM_Macros} 

\copyrightnotice{978-1-5386-5477-4/18/\$31.00 {\copyright}2018 IEEE}

\toappear{2018 IEEE International Workshop on Machine Learning for
  Signal Processing, Sept.\ 17--20, 2018, Aalborg, Denmark} 

\title{Dynamical Component Analysis (DyCA):\\
 Dimensionality reduction for high-dimensional deterministic time-series} 
\name{Bastian Seifert, Katharina Korn, Steffen Hartmann, Christian Uhl
  \thanks{This work is supported by the European Regional Development
    Fund (EFRE).}} 
\address{Ansbach University of Applied Sciences \\
  Faculty of Engineering Sciences \\
  Ansbach, Germany \\
bastian.seifert@hs-ansbach.de, christian.uhl@hs-ansbach.de}
\graphicspath{{pics/}}

\begin{document}
%

\maketitle
\begin{abstract}
    Multivariate signal processing is often based on dimensionality
    reduction techniques. We propose a new method, Dynamical Component
    Analysis (DyCA), leading to a classification of the underlying
    dynamics and - for a certain type of dynamics - to a signal
    subspace representing the dynamics of the data. In this paper the
    algorithm is derived leading to a generalized eigenvalue problem
    of correlation matrices. The application of the DyCA on
    high-dimensional chaotic signals is presented both for simulated
    data as well as real EEG data of epileptic seizures.
\end{abstract}
\section{Introduction}
\label{sec:intro}

Classic dimensionality reduction techniques, like principal component
analysis (PCA)~\cite{Pearson:1901} or independent component analysis
(ICA)~\cite{Hyvaerinen.Oja:2000a}, are widely used as a preprocessing
step in the analysis of multivariate time-series. PCA aims at
projections leading to the largest possible variances of the signal in
each direction, but the obtained temporal signals are not optimized to
describe the dynamics of the signal. ICA on the other hand relies on
the assumption that the time-series can be split in mutually
independent signals. There are other approaches like forecastable
component analysis (ForeCA)~\cite{Goerg:2013a} relying on
forecastiblity measure, approaches based on multivariate
autoregressive
models~\cite{Vidaurre.Rezek.Harrison.Smith.Woolrich:2014a}, or
approaches based on Granger causality~\cite{Kim:2012a}. An overview of
conventional technqiues is presented in~\cite{Pena.Poncela:2006a}.

As these techniques always rely on some sort of stochastic model
assumption, they are not very well suited for the dimensionality
reduction of multivariate time-series data with a strong deterministic
part. Reduction of dimensionality of multivariate time-series is e.g.\
relevant for signals sampled by more sensors than the dimensionality
of the underlying system. A typical example of such systems is the
electroencephalogram of epileptic seizures, where one has many sensors
but a very regular, low-dimensional behaviour of the measured system.
The dimensionality reduction technique we introduce relies on a
special deterministic model assumption, suitable for example for the
reduction of some chaotic time-series. The proposed method is quite
similar to the methods of principal interacting and principal
oscillation patterns (PIPs and POPs)~\cite{Hasselmann:1988a} used in
geophysical sciences. In some sense the method we are presenting can
be interpreted as a generalization of the PIPs and POPs method.
Furthermore there are some technical similarities with methods for
transfer operator approximation based on delay coordinates, which are
applied in fluid or molecular
dynamics~\cite{Klus.Nueske.Koltai.Wu.Kevrekidis.Schuette.Noe:2018a}.

Chaotic time-series forecasting by reservoir computing has recently
resulted in very interesting results, outperforming all tools
available up to
now~\cite{Pathak.Lu.Hunt.Girvan.Ott:2017a,Antonik.Gulina.Pauwels.Massar:2018a,Pathak.Hunt.Girvan.Lu.Ott:2018a}.
We suggest the proposed dimensionality reduction technique as an
adequate preprocessing step for reservoir computing of
high-dimensional spatio-temporal data.

The structure of the paper is as follows. First we derive Dynamical
Component Analysis (DyCA) using variational calculus. It is shown that
DyCA corresponds to a generalized eigenvalue problem. The eigenvalues
of the generalized eigenvalue problem tell the quality of a fit of the
data to a system of ordinary differential equations of special form.
The application of DyCA to high-dimensional simulated data based on
the Rössler system is presented in Section~\ref{sec:Roessler}. In
Section~\ref{sec:ApplicationEEG} EEG data of epileptic seizures
expecting Shilnikov chaos are investigated by the proposed method.

\section{Dynamical Component Analysis}%
\label{sec:DyCA}%

Let $q(t) \in \mathbb{R}^N$ be a multivariate time-series with its
dynamics being described by a low-dimensional system of ordinary
differential equations. I.e., we can decompose the signal in time-dependent
amplitudes $x_i(t)$ and vectors $w_i \in \mathbb{R}^N$,
\begin{equation}
\label{eq:SigDecomp}
q(t) = \sum_{i=1}^n x_i(t) w_i,
\end{equation}
with the dynamics of the amplitudes described by the set of differential equations
\begin{equation}
    \label{eq:LinearPartODE}
    \begin{split}        
	\dot x_1 & = \sum_{k=1}^n a_{1,k} x_k \\  
        & \vdots  \\
        \dot x_m &= \sum_{k=1}^n a_{m,k} x_k \\
    \end{split}
\end{equation}
and
\begin{equation}
    \label{eq:NonlinearPartODE}
    \begin{split}
        \dot x_{m+1} &= f_{m+1}(x_1, x_2, \ldots, x_n) \\  
        & \vdots  \\
        \dot x_n &= f_n (x_1, x_2, \ldots, x_n),
    \end{split}    
\end{equation}
where $n \ll N$ and $f_j$ are non-linear smooth functions. We
assume that we neither know the parameters $a_{i,k}$ nor the exact
form of the functions $f$.

To generate projection vectors $u_i, v_j \in \mathbb{R}^N$ approximating
the above mentioned amplitudes $x_i(t)$ we minimize
the least square error cost function
\begin{equation}
    \label{eq:LeastSquareCostFunction}%
    D(u,v,a) = \frac{\timeavg{ \norm{ \dot{q}^\top u - \sum_j a_j q^\top
             v_j }_2^2 }} {\timeavg{ \norm{\dot{q}^\top u}_2^2}}  
\end{equation}
where $\timeavg{\argument}$ denotes the time average. Denote the
correlation matrices of the signal with itself, of the signal with its
derivatives, and the signal derivatives with itself by
$C_0 = \timeavg{q q^\top}, C_1 = \timeavg{\dot{q} q^\top},$ and
$C_2 = \timeavg{\dot{q} \dot{q}^\top}$, respectively. Then we can
rewrite the cost function as
\begin{equation}
    \label{eq:LeastSquareCostFunctioRewritten}
    \begin{split}
        &D(u,v,a) \\
        &=  \frac{\timeavg{ \norm{ \dot{q}^\top u - \sum_j a_j q^\top
              v_j }_2^2 }} {\timeavg{ \norm{\dot{q}^\top u}_2^2}} \\
        &=  \frac{\timeavg{ (\dot{q}^\top u - \sum_j a_j q^\top
            v_j)^\top  (\dot{q}^\top u - \sum_j a_j q^\top
            v_j)}} {\timeavg{(\dot{q}^\top u)^\top
            (\dot{q}^\top u)}} \\ 
        &=  \frac{(u^\top C_2 u) - 2 \sum_j a_j (u^\top C_1 v_j) +
          \sum_{j,k} a_j a_k (v_j^\top C_0 v_k)}{u^\top C_2 u} \\
        &= 1 - 2 \sum_j a_j \frac{u^\top C_1 v_j}{u^\top C_2 u} +
        \sum_{j,k} a_j a_k \frac{ v_j^\top C_0 v_k}{u^\top C_2 u}.
    \end{split}
\end{equation}
The minimum of $D$ can be analytically calculated using variation with
respect to each variable. For this the partial derivatives with
respect to the variables $u,v,$ and $a$ are derived and their minima
determined.

The partial derivative with respect to $u^\top$ is
\begin{equation}
    \label{eq:DerivativeWithRespectToU}
    \begin{split}
        \frac{\partial D}{\partial u^\top} 
        &= - 2 \sum_j a_j \frac{C_1 v_j (u^\top C_2 u) - 2(u^\top C_1
          v_j) C_2 u}{(u^\top C_2 u)^2} \\
        &\quad - 2 \sum_{j,k} a_j a_k
        \frac{(v_j^\top C_0 v_k) C_2 u}{(u^T C_2 u)^2}.            
    \end{split}
\end{equation} 
Setting the derivative to zero leads to
\begin{equation}
    \label{eq:EigenvaluesByULongVersion}
    \begin{split}
        &(2 \sum_j a_j (u^\top C_1 v_j) - \sum_{j,k} a_j a_k (v_j^\top C_0
        v_k)) C_2 u  \\
        &= (u^\top C_2 u) \sum_j a_j C_1 v_j.        
    \end{split}
\end{equation}
Let $\mu = 2(\sum_j a_j (u^\top C_1 v_j) - \sum_{jk} a_j a_k
(v_j^\top C_0 v_k))$ and 
$\tau = u^\top C_2 u$
then \eqref{eq:EigenvaluesByULongVersion} reads
\begin{equation}
    \label{eq:EigenvaluesByU}
    \mu C_2 u = \tau C_1 \sum_j a_j v_j. 
\end{equation}

The partial derivative with respect to $v_r$ is
\begin{equation}
    \label{eq:DerivativeWithRespecToV}
    \frac{\partial D}{\partial v_r} = 
    - 2 a_r \frac{u^\top C_1}{(u^\top C_2 u)^2}
    + 2 a_r \sum_j a_j \frac{v_j^\top C_0}{(u^\top C_2 u)^2} 
\end{equation}
For $\frac{\partial D}{\partial v_r} = 0$ we therefore obtain
\begin{equation}
    \label{eq:EigenvaluesByV}
    \vect{u}^\top C_1 = (\sum_j a_j v_j^\top) C_0. 
\end{equation}

Calculating the partial derivative with respect to $a_r$ results in 
\begin{equation}
    \label{eq:DerivativeWithRespectToA}
    \frac{\partial D}{\partial a_r} = - 2 \frac{u^\top C_1
      v_r}{u^\top C_2 u} 
    + 2 \sum_j a_j \frac{v_j^\top C_0 v_r}{u^\top C_2 u}  
\end{equation}
and setting the derivative to 0 leads to 
\begin{equation}
\label{eq:SolutionA}
u^\top C_1 v_r  = \sum_j v_j^\top C_0 v_r.
\end{equation}
Note that both multiplying \eqref{eq:EigenvaluesByU} from left with
$u^\top$ and \eqref{eq:EigenvaluesByV} from right with $v$ lead to
\eqref{eq:SolutionA} proving the consistency of the calculation.

Assuming the existence of the inverse $C_0^{-1}$ of the correlation
matrix $C_0$, \eqref{eq:EigenvaluesByV} can be rewritten as
\begin{equation}
\label{eq:EigenvaluesByVRewritten}
\sum_j a_j \vect{v}_j = C_0^{-1} C_1^\top \vect{u}.
\end{equation}
Inserting \eqref{eq:EigenvaluesByVRewritten} into
\eqref{eq:EigenvaluesByU} a generalized eigenvalue 
problem is obtained  
\begin{equation}
\label{eq:GeneralizedEigenvaluedProblem}
C_1 C_0^{-1} C_1^\top \vect{u}  = \lambda C_2 \vect{u},
\end{equation}
where $\lambda = \frac{\mu}{\tau}$.

Inserting \eqref{eq:EigenvaluesByU} and \eqref{eq:EigenvaluesByV} into
\eqref{eq:LeastSquareCostFunctioRewritten} yields
\begin{equation}
    \label{eq:2}
    \begin{split}
	D_{min}
	& = 1 - 2 \sum_j a_j  \frac{u^\top C_1 v_j}{u^\top C_2 u} 
	+ \sum_{j,k} a_j a_k \frac{v_j^\top C_0 v_k}{u^\top C_2 u} \\ 
	& = 1 - \frac{2 \sum_j a_j u^\top C_1 v_j +
          \sum_{j,k} a_j a_k v_j^\top C_0 v_k}{u^\top C_2 u} \\ 
	& = 1 - \frac{\mu}{\tau} \\
	& = 1 - \lambda .      
    \end{split}
\end{equation}
That means, similiar to principal component analysis (PCA), the
eigenvalues $\lambda_i$ of the generalized eigenvalue problem
(eq.\eqref{eq:GeneralizedEigenvaluedProblem}) indicate the quality of
the least-square-fit of the linear differential
equations~\eqref{eq:LinearPartODE}. The eigenvalue spectrum allows for
an identification of amplitudes interacting as
$\dot x_i(t) = \sum_j a_{i,j} x_j(t)$ by projecting the signal $q(t)$
onto the corresponding eigenvector $u$, i.e. $x_i(t) = q(t)^\top u_i$.
By choosing an appropriate threshold one can obtain a projection
subspace spanned by the $m$ corresponding eigenvectors $u_i$.
Calculation of $C_2 u_i$ by \eqref{eq:EigenvaluesByU} leads to a set
of $m$ vectors (as linear combination of the unknown vectors $v_i$)
which then span another $m$ dimensional subspace. The span of these
both $m$-dimensional subspaces
\begin{equation}
\label{eq:ApproximationSubspace}
\mathsf{span} \{ u_1, \ldots , u_m, C_1^{-1} C_2 u _1, \ldots ,
C_1^{-1} C_2 u_m \} = \mathbb{R}^n  
\end{equation}
approximates the complete $n$-dimensional subspace in which the system
evolution can be described by a set of differential equations
(eq.\eqref{eq:LinearPartODE} and \eqref{eq:NonlinearPartODE}) if $m$
is not too small. Obviously, if $m < n/2$ this would not work.

Note that the size of the matrices $C_0, C_1$ and $C_2$ is
$N \times N$, which is small compared to the length of a typical
time-series. Hence the application of DyCA as a preprocessing step is
computationally cheap. The invertibility of $C_0$ relies on the
different sensors measuring independent signals. In most
applications this is the case due to inherent measuring noise.

\section{Application to the Rössler system}
\label{sec:Roessler}%

The Rössler attractor~\cite{Roessler:1976a} is a strange attractor
given by the system of ordinary differential equations
\begin{equation}
    \label{eq:RoesslerSystem}
    \begin{split}
        \dot{x_1} &= -x_2 - x_3 \\
        \dot{x_2} &= x_1 + a x_2\\
        \dot{x_3} &= b - c x_3 + x_1 x_2,
    \end{split}
\end{equation}
with $a = 0.15, b = 0.2,$ and $c = 10$. For the application of DyCA a
trajectory of this system was obtained using a $(4,5)$-Runge-Kutta
integration method. Then the data was embedded in a $25$-dimensional
space with additional multiplicative Gaussian noise. We used an
exemplary signal to noise ratio of $15 \mathsf{dB}$. As definition of
signal to noise ratio we rely on the formula
$SNR = \tfrac{A}{\sigma}$, where $A$ is the signal mean and $\sigma$
the standard deviation of the noise.

\begin{figure}
    \centering
    \includegraphics[width=0.43\textwidth]{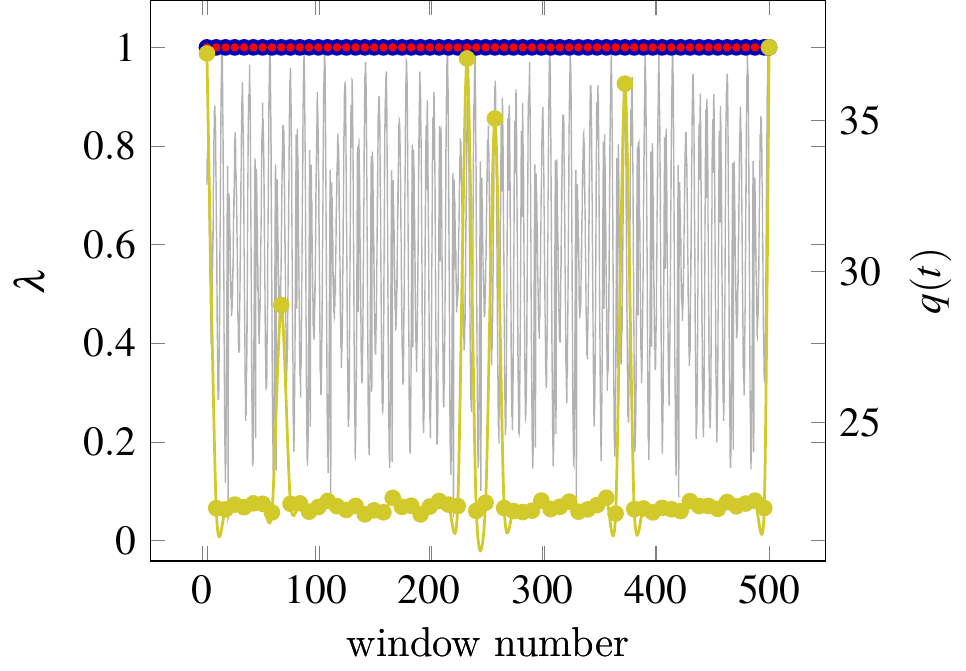}
    \caption{The three largest generalized eigenvalues for each window
      colored blue, red and yellow, respectively. The first two are
      always equal to one, while the third one drops close to zero except
      for some exceptional parts. In the background one time-series of
    the high-dimensional signal is shown.}
    \label{fig:EigenvaluesRoessler}
\end{figure}

\begin{figure}
    \centering
    \includegraphics[width=0.43\textwidth]{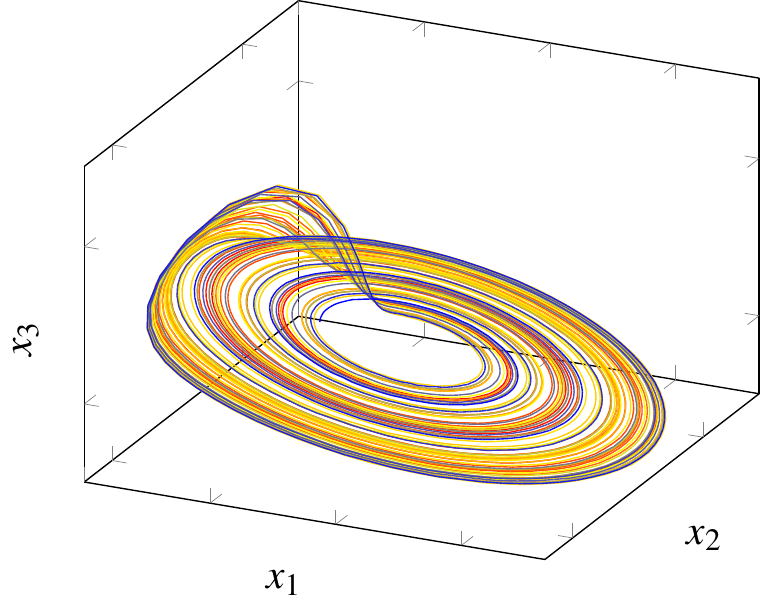}
    \caption{The projection of the 25-dimensional time-series in
      phase-space using DyCA. The color indicates the time evolution.}
    \label{fig:RoesslerProjected}
\end{figure}

\begin{figure}
    \centering
    \includegraphics[width=0.43\textwidth]{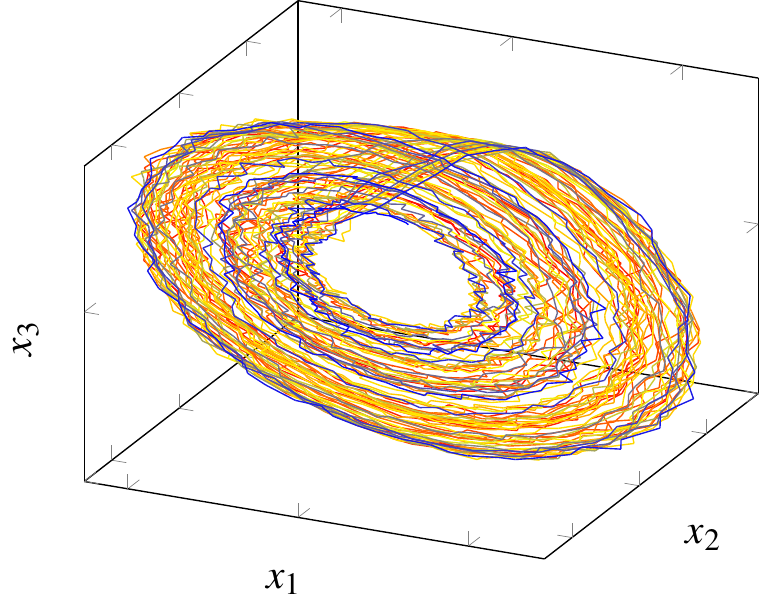}    
    \caption{The projection of the 25-dimensional time-series in
      phase-space using PCA. The color indicates
      the time evolution.}
    \label{fig:RoesslerPCA}
\end{figure}

\begin{figure}
    \centering
    \includegraphics[width=0.43\textwidth]{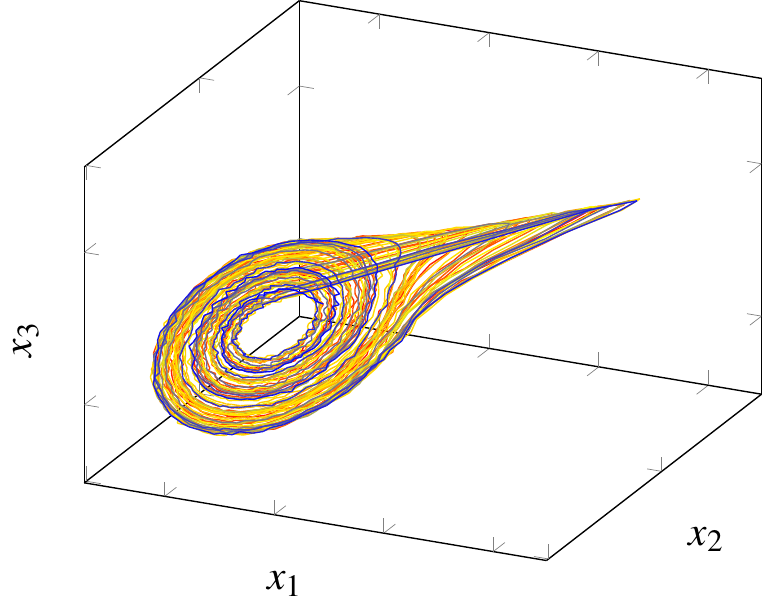}    
    \caption{The projection of the 25-dimensional time-series in
      phase-space using ICA. The color indicates the time evolution.}
    \label{fig:RoesslerICA}
\end{figure}

The generalized eigenvalue spectrum of DyCA applied to
the 25-dimensional simulated data is illustrated in Fig.~\ref{fig:EigenvaluesRoessler}.
As expected, according to \eqref{eq:RoesslerSystem},
the two largest eigenvalues are equal to one due to the two linear equations
in \eqref{eq:RoesslerSystem}. The third eigenvalue is evidently below one
according to one nonlinear equation in \eqref{eq:RoesslerSystem}. 
The span of the projection vectors
$\mathsf{span}\{u_1,u_2, C_2 u_1, C_2 u_2\}$ is, with respect to
numerical tolerances, of dimension $3$. Projecting with the projection
vectors $u_1,u_2$ and $v_2 = C_2 u_2$ leads to the phase-potrait
illustrated in Fig.~\ref{fig:RoesslerProjected}. In
Fig.~\ref{fig:RoesslerPCA} and \ref{fig:RoesslerICA} the
phase-potraits of the data obtained by dimensionality reduction using
PCA and ICA are shown. Subjective comparison of the obtained figures
suggests that the inherent dynamics of the data is more accurately
represented and the noise is reduced in a larger amount in the data
projected with DyCA than in the data projected with PCA or ICA. The
projections obtained by PCA resemble the results one would obtain by
picking three time-series out of the twenty-five of the original
multivariate signal at random.

\section{Application to epileptic EEG data}
\label{sec:ApplicationEEG}%

A typical example where the assumptions \eqref{eq:LinearPartODE} and
\eqref{eq:NonlinearPartODE} are fulfilled is the EEG data of an
epileptic seizure. This is due to the conjectured appearance of
Shilnikov chaos in epileptic seizures. Using bifurcation analysis the
existence of Shilnikov chaos in various theoretical models was shown
by van Veen and Liley~\cite{vanVeen.Liley:2006}. In
\cite{Friedrich.Uhl:1996a} a system of ordinary differential equations
of the form
\begin{equation}
    \label{eq:ShilnikovSystemEEG}
    \begin{split}
        \dot{x_1} &= x_2 \\
        \dot{x_2} &= x_3 \\
        \dot{x_3} &= f(x_1,x_2,x_3),
    \end{split}
\end{equation}
with $f$ being a non-linear polynomial function, was assumed to model
epileptic encephalograms. Since this model relies on two linear and
one non-linear equations, we assume that the conditions on the
applicability of DyCA are fulfilled.

\begin{figure}
    \centering
    \includegraphics[width=0.43\textwidth]{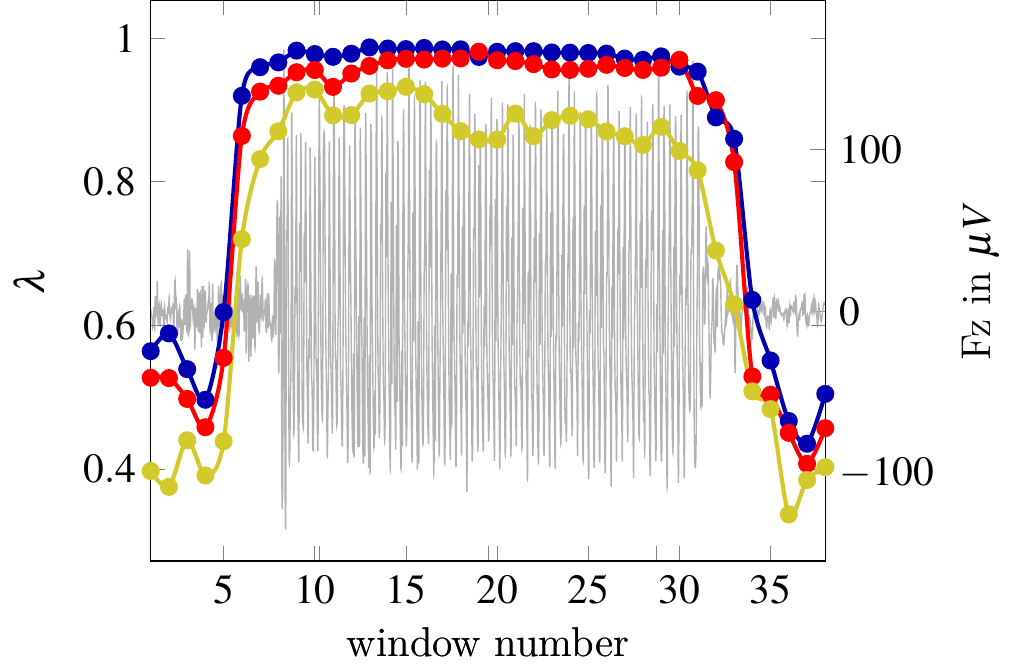}
    \caption{The three largest generalized eigenvalues of DyCA for
      each window colored blue, red and yellow, respectively. In the
      background the Fz electrode of the EEG is shown.}
    \label{fig:Eigenvalues}
\end{figure}

As data we considered a set of EEG data containing stages before,
after and during an epileptic seizure. The data was sampled using 25
sensors with $256$ Hertz sample rate. The signal to noise ratio is
approximately $16 \mathsf{dB}$. As preprocessing step the data was
bandpass filtered with a zerophase filter with cut-off frequencies of
$0.5$ and $30$ Hertz. The data was partitioned in windows of one
second length. Then DyCA was applied on each window. As can be seen in
Fig.~\ref{fig:Eigenvalues} the assumption of a system of the
form~\eqref{eq:ShilnikovSystemEEG} during an epileptic seizure can be
accepted, since the two largest eigenvalues are nearly $1$ during the
absence.

\begin{figure}
    \centering
    \includegraphics[width=0.43\textwidth]{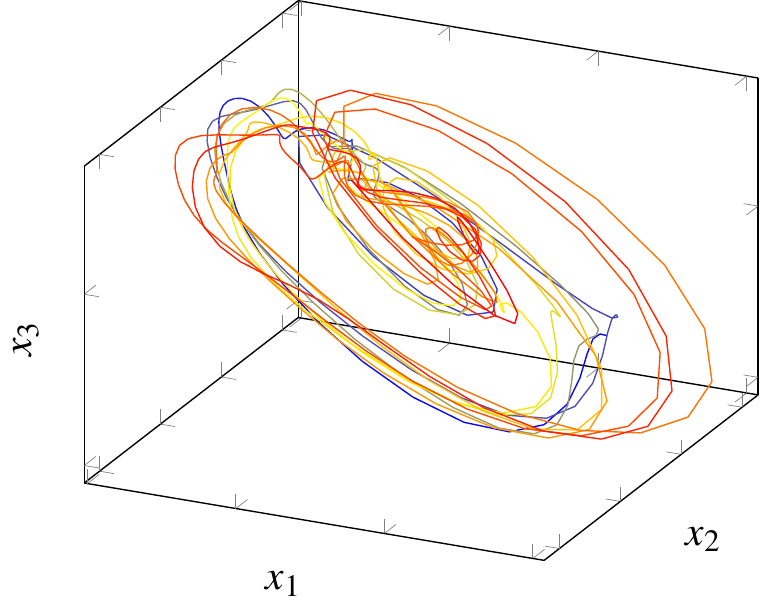}
    \caption{Projected EEG time-series in phase-space using the
      projection obtained by DyCA on the dataset.}
    \label{fig:ShilnikovExampleWindow}
\end{figure}

\begin{figure}
    \centering
    \includegraphics[width=0.43\textwidth]{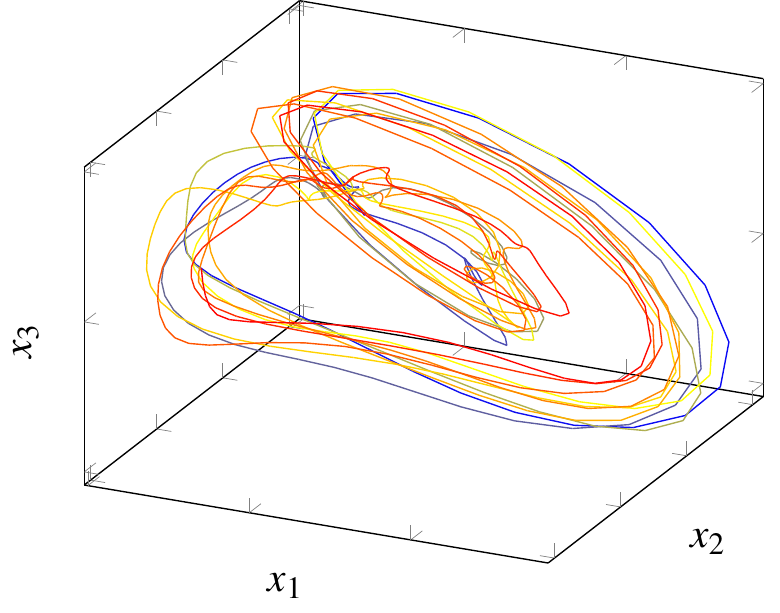}
    \caption{Projected EEG time-series in phase-space using the
      projection from Fig.~\ref{fig:ShilnikovExampleWindow} on another
      window.} 
    \label{fig:ShilnikovExampleWhole}
\end{figure}

Since DyCA is proposed as a preprocessing method for machine learning
applications, we need to show that the projection calculated on one
window is able to represent other parts of the time-series, as well.
Fig.~\ref{fig:ShilnikovExampleWindow} and
\ref{fig:ShilnikovExampleWhole} show that if one uses the projection
obtained on one window of the data to project another window, the
underlying dynamics is still preserved.
Hence, if the applicability assumptions are fulfilled, DyCA is
suitable as preprocessing tool for analysis of high-dimensional
deterministic time-series.

\section{Discussion and Conclusion}
\label{sec:Discussion}%

We conclude that DyCA is a suitable tool for dimensionality reduction
of high-dimensional time-series, provided the underlying dynamics can
be described by a system of ordinary differential equations of the
form \eqref{eq:LinearPartODE} and \eqref{eq:NonlinearPartODE}. It has
been shown that DyCA can get rid of noise more efficiently than PCA
and ICA. Furthermore DyCA is able to preserve the dynamics of
spike-waves in epileptic EEG data. Since the calculation of the
projection matrices of DyCA is simply solving a generalized eigenvalue
problem, the procedure is computationally cheap. Hence it is suggested
to establish DyCA as a more reliable alternative to PCA as
preprocessing step in the analysis of multivariate deterministic
time-series.

Further studies are needed to show that DyCA improves the prediction
ability of reservoir computing approaches. These will be conducted in
future works.


\bibliographystyle{IEEEbib}
\bibliography{DyCAMLSP}
\vfill
\pagebreak
\end{document}

%% file: DSBM_Macros.tex
\renewcommand{\mathbb}[1]{\mathbbm{#1}}

\allowdisplaybreaks

\let\originalleft\left
\let\originalright\right
\renewcommand{\left}{\mathopen{}\mathclose\bgroup\originalleft}
\renewcommand{\right}{\aftergroup\egroup\originalright}

\newcommand{\argument}       {\,\cdot\,}

\newcommand{\vect}[1]{\boldsymbol{#1}}
\DeclarePairedDelimiter{\timeavg}{\langle}{\rangle_t}

\DeclarePairedDelimiter{\norm}{\lVert}{\rVert}

